\documentclass[a4paper]{spie}  

\usepackage{amsmath,amsfonts,amssymb}
\usepackage{graphicx}
\usepackage[colorlinks=true, allcolors=blue]{hyperref}
\usepackage{upgreek}

\title{Soft proton experiments supporting the development of astronomical X-ray instrumentation}

\author[a]{S.\,J.~Diebold}
\author[a]{B.~Heß}
\author[a]{F.~Pfeifle}
\author[b]{L.\,Ph.\,H.~Schmidt}
\author[a]{C.~Tenzer}
\author[a]{T.~Wildfang}
\author[c]{D.~Ferreira}
\author[d]{M.~Freyberg}
\author[c]{S.~Massahi}
\author[c]{D.~Paredes-Sanz}
\author[a]{E.~Perinati}
\author[a]{A.~Santangelo}
\author[a]{B.~Stelzer}
\author[c]{S.~Svendsen}

\affil[a]{Institut für Astronomie und Astrophysik, Eberhard Karls Universität Tübingen, Sand 1, 72076 Tübingen, Germany}
\affil[b]{Institut für Kernphysik, Goethe-Universität Frankfurt, Max-von-Laue-Str.\@ 1, 60438~Frankfurt~am~Main, Germany}
\affil[c]{Department of Space Research and Technology, Technical University of Denmark, Elektrovej, Bygning 328, 2800 Kgs. Lyngby, Denmark}
\affil[d]{Max-Planck-Institut für extraterrestrische Physik, Gießenbachstraße~1, 85748~Garching, Germany}

\authorinfo{Corresponding author: Sebastian Diebold\\E-mail: diebold@astro.uni-tuebingen.de, Telephone: +49 7071 29 78604}

\begin{document} 
\maketitle

\begin{abstract}
Orbital soft protons can severely degrade the performance of astronomical X-ray observatories. On the one hand, they may cause permanent radiation damage to X-ray detectors; on the other hand, they introduce an irreducible background component.

Our low-energy grazing-incidence scattering setup provides experimental data for the validation of radiation transport simulations used in the assessment of future X-ray missions. Recent measurements indicate that a significant fraction of protons scattered from X-ray optics undergo charge exchange and therefore cannot be mitigated by magnetic diverters. In addition, we present initial proton-transmission measurements through a thin X-ray filter and compare them qualitatively with TRIM simulations. Quantitative measurements of energy loss and charge-exchange fractions, together with comparisons with TRIM and Geant4, are planned.

In this publication, we present the upgraded experimental setup together with commissioning measurements for grazing-incidence scattering and proton transmission through thin X-ray filters.
\end{abstract}



\section{INTRODUCTION}
\label{sec:intro}

Since the rapid in-orbit degradation of the front-illuminated CCDs aboard the \textit{Chandra X-ray Observatory}\cite{Weisskopf2003} shortly after launch, soft protons have become a major concern in the design and assessment of astronomical X-ray missions. Their impact is twofold: they may degrade the performance of X-ray detectors and they contribute to the instrumental background\cite{Campana2024}. While the radiation hardness of modern X-ray sensors has improved considerably over the past decades, such that radiation damage is generally expected to remain acceptable over typical mission lifetimes, the soft-proton-induced background continues to pose a significant challenge. This issue becomes even more critical for future observatories with large effective area such as \textit{NewAthena} \cite{Barret2020}. Unlike energetic charged particles, soft protons are focused by grazing-incidence X-ray optics onto the focal plane, where they often produce signals that are indistinguishable from X-ray events, lie within the scientifically relevant energy band, and, therefore, cannot be removed during data analysis.

The first dedicated soft proton scattering experiments were initiated in the few months after the radiation damage on \textit{Chandra} had been identified and shortly before the launch of \textit{XMM-Newton} at the end of 1999\cite{Rasmussen1999}. Around the same time, the Institut für Astronomie und Astrophysik (IAAT) of the University of Tübingen performed the first irradiation experiments on EPIC pn-CCDs developed for \textit{XMM-Newton} by the Max Planck Semiconductor Laboratory\footnote{\url{https://www.hll.mpg.de/}}\cite{Kendziorra2000}. The experimental facility was subsequently upgraded for grazing-incidence scattering measurements on \textit{eROSITA} mirror samples \cite{Diebold2015,Diebold2017}. Following the final shutdown of the accelerator facility in Tübingen in 2018, the setup was transferred to the Institut für Kernphysik Frankfurt (IKF) of the Goethe University in Frankfurt am Main, where it was primarily used to investigate silicon pore optics (SPO) samples developed for \textit{NewAthena} \cite{Amato2021}.

\begin{figure}[ht]
\centering
\includegraphics[width=.8\textwidth]{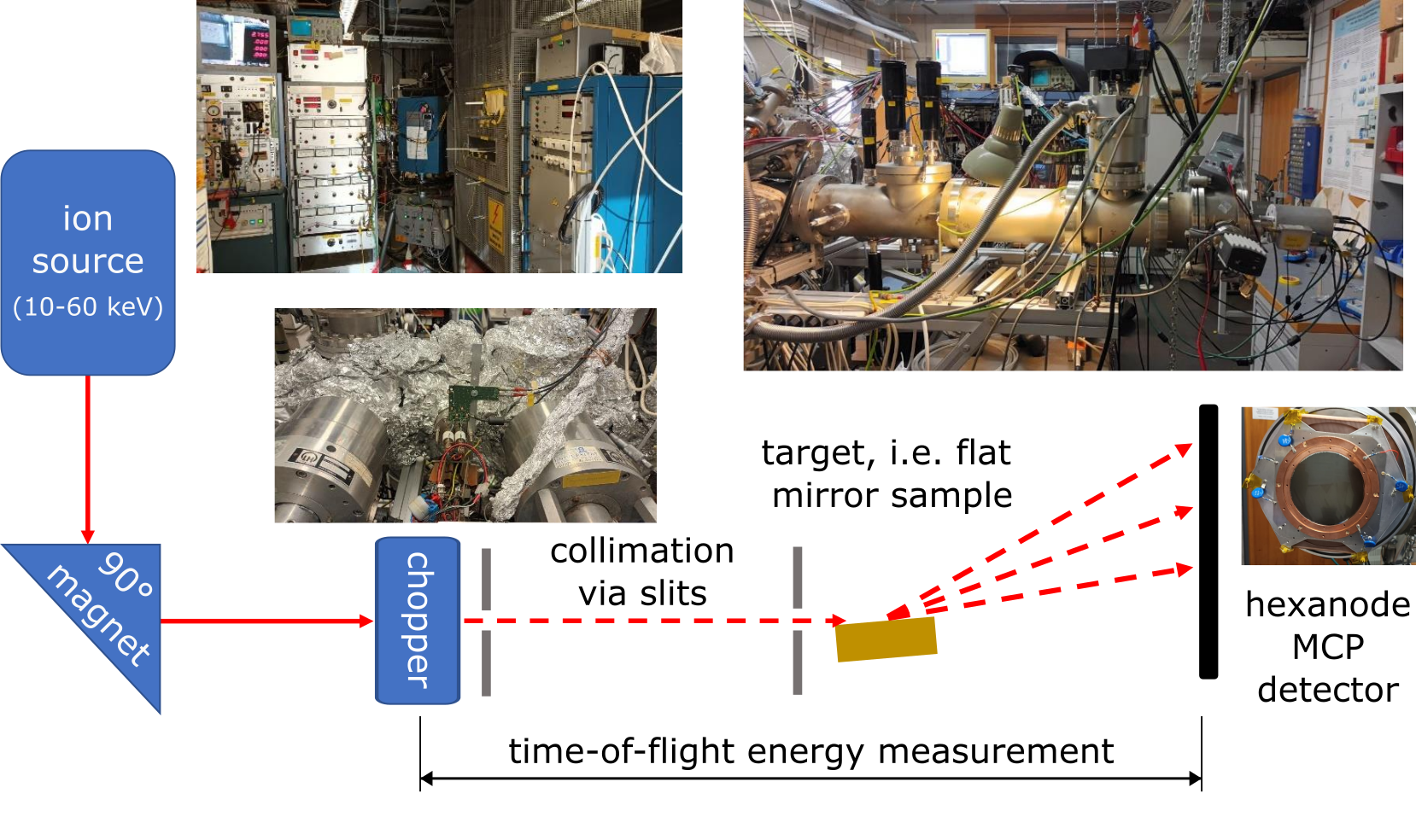}
\caption{Schematic overview with photographs of the low-energy scattering setup at the ion implanter at IKF. An electrostatic ion source generates the proton beam, which is cleaned by a $90^\circ$ bending magnet. A beam chopper reduces the flux to individual protons for time-of-flight measurements before the beam is collimated by two pairs of slits. After scattering from the target, the proton distribution is recorded by a position-sensitive hexanode MCP detector. Figure adapted from Diebold et al.\ (2024) \cite{Diebold2024}.}
\label{fig:setup}
\end{figure}

As it proved difficult to reproduce the measurement precision previously achieved in Tübingen, and because scientific interest increasingly shifted towards proton energies below 100\,keV, which is outside the operating range of a Van de Graaff accelerator, a new scattering experiment was established at the ion implanter of IKF. The facility covers an energy range of approximately 10--60\,keV, and scattered particles are detected on a position-sensitive large-area microchannel plate (MCP) detector equipped with a hexanode delay-line readout\footnote{\url{https://www.roentdek.com/info/Delay_Line/}}. In combination with a beam chopper, the nanosecond timing capability of the detector enables reconstruction of the proton energy by time-of-flight (ToF) measurements. Figure~\ref{fig:setup} provides an overview of the experimental setup. A detailed description of the facility and a review of the historical development of soft proton experiments for X-ray instrumentation are given by Diebold et al.\ (2024) \cite{Diebold2024}.

In this publication we summarize the recent progress with the setup at the IKF implanter and presents corresponding test and commissioning measurements. Section~\ref{sec:mount} briefly sketches an observed discrepancy in measurements with an SPO target, and describes the upgraded target mount, the newly implemented methods for determining the proton grazing angle, and presents a first beam test with a flat multilayer target. Section~\ref{sec:trans} presents the first proton transmission measurements using an \textit{eROSITA} flight spare filter, together with the experimental challenges encountered and the mitigation measures that were implemented. Finally, Section~\ref{sec:concl} summarizes the conclusions and outlines planned measurement campaigns.


\section{TARGET MOUNT AND DETERMINATION OF GRAZING ANGLE}
\label{sec:mount}

Previous measurements at the Van de Graaff facility at IKF focused on grazing-incidence scattering from silicon pore optics (SPO) samples. Preliminary measurements with the new setup at the implanter facility at IKF revealed a systematic deviation from the predictions of the Remizovich model\cite{Remizovich1980}. At grazing angles of about $0.5^\circ$ and above, the peak of the forward scattering distribution occurred at approximately 1.85 times the grazing angle, in agreement with the Remizovich model and close to the specular reflection angle. At smaller grazing angles, however, the peak shifted systematically toward larger scattering angles. This behavior was consistently observed at different beam energies and motivated a detailed investigation of the grazing-angle determination.

\begin{figure}[ht]
\centering
\includegraphics[width=.5\textwidth]{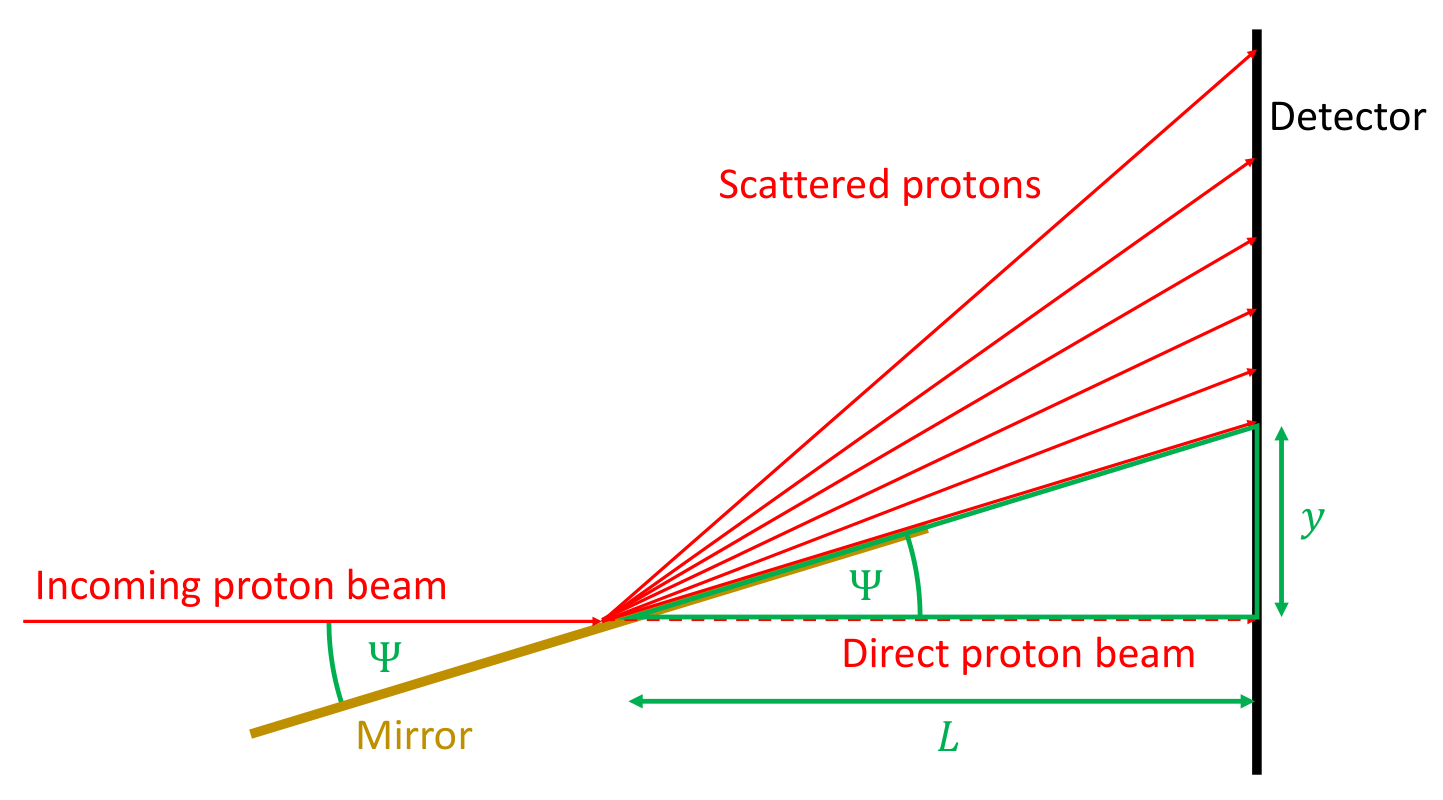}
\caption{Principle of the geometric determination of the grazing angle $\Psi$. For a flat target, the lower edge of the scattering distribution corresponds to the projection of the target surface onto the detector plane and therefore provides an independent estimate of the grazing angle.}
\label{fig:shadow}
\end{figure}

\begin{figure}[ht]
\centering
\includegraphics[width=.9\textwidth]{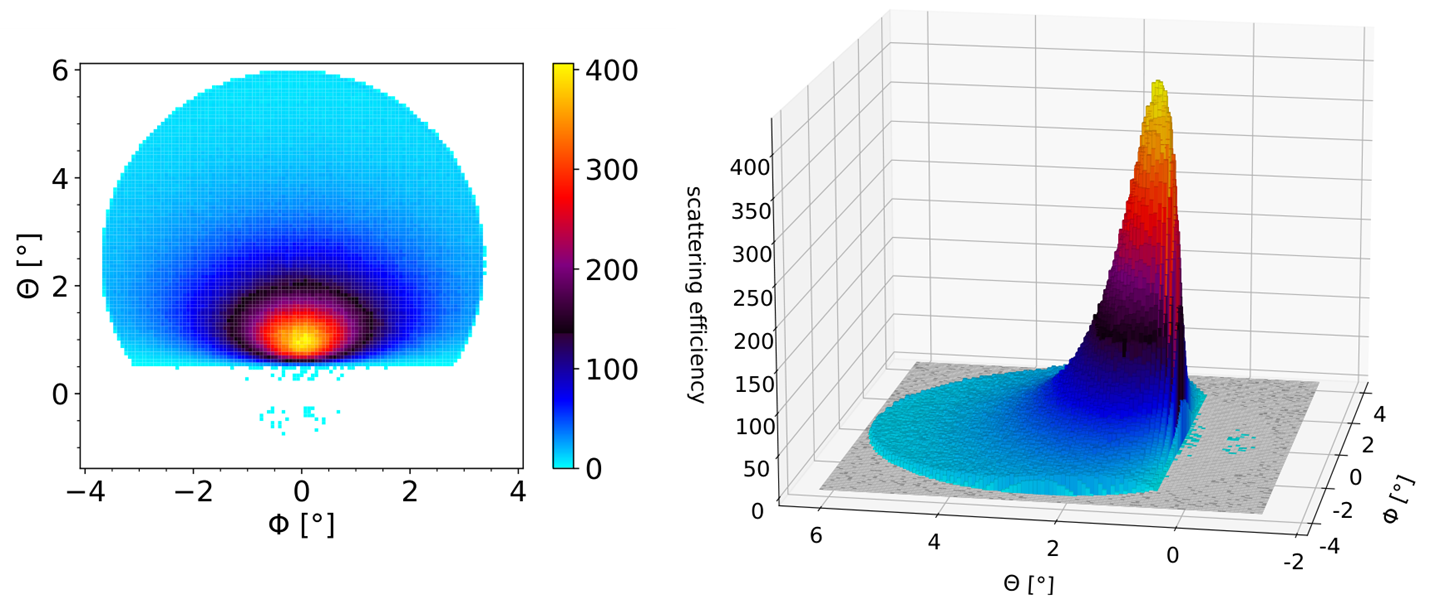}
\caption{Preliminary scattering efficiency measurement obtained for an SPO sample at an incident proton energy of 35\,keV and a grazing angle of approximately $0.5^\circ$. Figure adapted from Diebold et al.\ (2024) \cite{Diebold2024}.}
\label{fig:scattering}
\end{figure}

Unlike the previous Van de Graaff setup, the grazing angle was determined geometrically from the shadow cast by the front and rear edges of the target and the corresponding readings of the linear manipulators supporting the target mount. An independent estimate was obtained from the shadow of the entire target surface, which appears as the lower boundary of the scattering distribution. The principle is illustrated schematically in Figure~\ref{fig:shadow}, while Figure~\ref{fig:scattering} shows an example scattering distribution.

The observed shift is compatible with earlier measurements on \textit{eROSITA} mirror samples, but it was not statistically significant due to lower angular resolution and limited statistics\cite{Diebold2015,Diebold2017}. Several possible explanations for it were investigated. One initial hypothesis considered an image-charge interaction between the incident proton and the metallic mirror coating. As the proton travels almost parallel to the surface, the induced image-charge could increase the effective grazing angle, resulting in an apparent shift of the scattering distribution. Since the interaction becomes stronger at smaller distances from the surface, such an effect would be expected to increase with decreasing grazing angle. Analytical estimates and numerical calculations, however, showed that the expected deflection is far below the experimental sensitivity and, therefore, cannot account for the observed discrepancy.

The most plausible explanation yet was eventually identified as the intrinsic curvature of the silicon plates used for the SPO samples, as discussed in the following section.


\subsection{Silicon pore optics sample curvature}
\label{ssec:SPO}

\begin{figure}[ht]
\centering
\begin{tabular}{lcr}
\includegraphics[width=.4\textwidth]{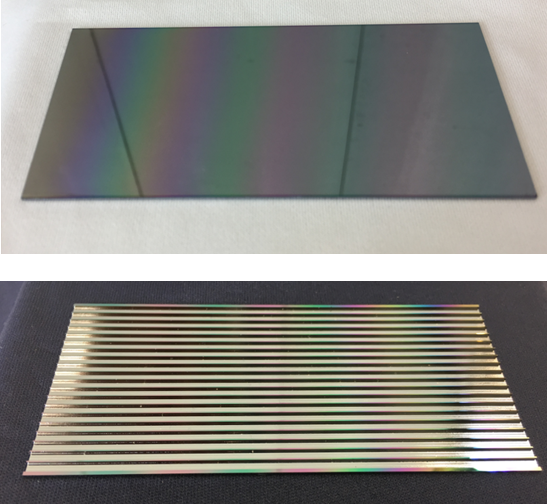} & \hspace{1cm} &
\includegraphics[width=.295\textwidth]{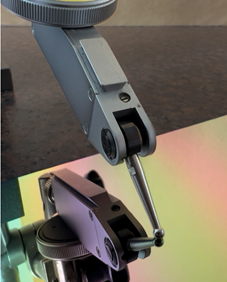}
\end{tabular}
\caption{Left: Photograph of coated top side and ribbed bottom side of the SPO sample previously measured. Figure adapted from Amato et al.\ (2021)\cite{Amato2021}. Right: Mahr MarTest 800 S micrometer gauge used for the mechanical surface measurements.}
\label{fig:SPO}
\end{figure}

The SPO samples used in the first measurements at the implanter setup had previously been investigated at the Van de Graaff facility. They were supplied by cosine research B.V.\footnote{\url{https://www.cosine.eu/}} and consist of 110\,mm long, 0.775\,mm thick silicon substrates coated with 10\,nm of Ir and 7\,nm of SiC. A photograph of the coated front surface and the ribbed rear side is shown in Figure~\ref{fig:SPO}.

In order to characterize the surface profile the target surface was measured mechanically. Prior finite-element simulations performed in the CAD software SolidWorks confirmed that the contact force of the measurement gauge (less than 150\,mN) produces surface deformations that are limited to a few nanometers and, therefore, are negligible.

\begin{figure}[ht]
\centering
\includegraphics[width=.85\textwidth]{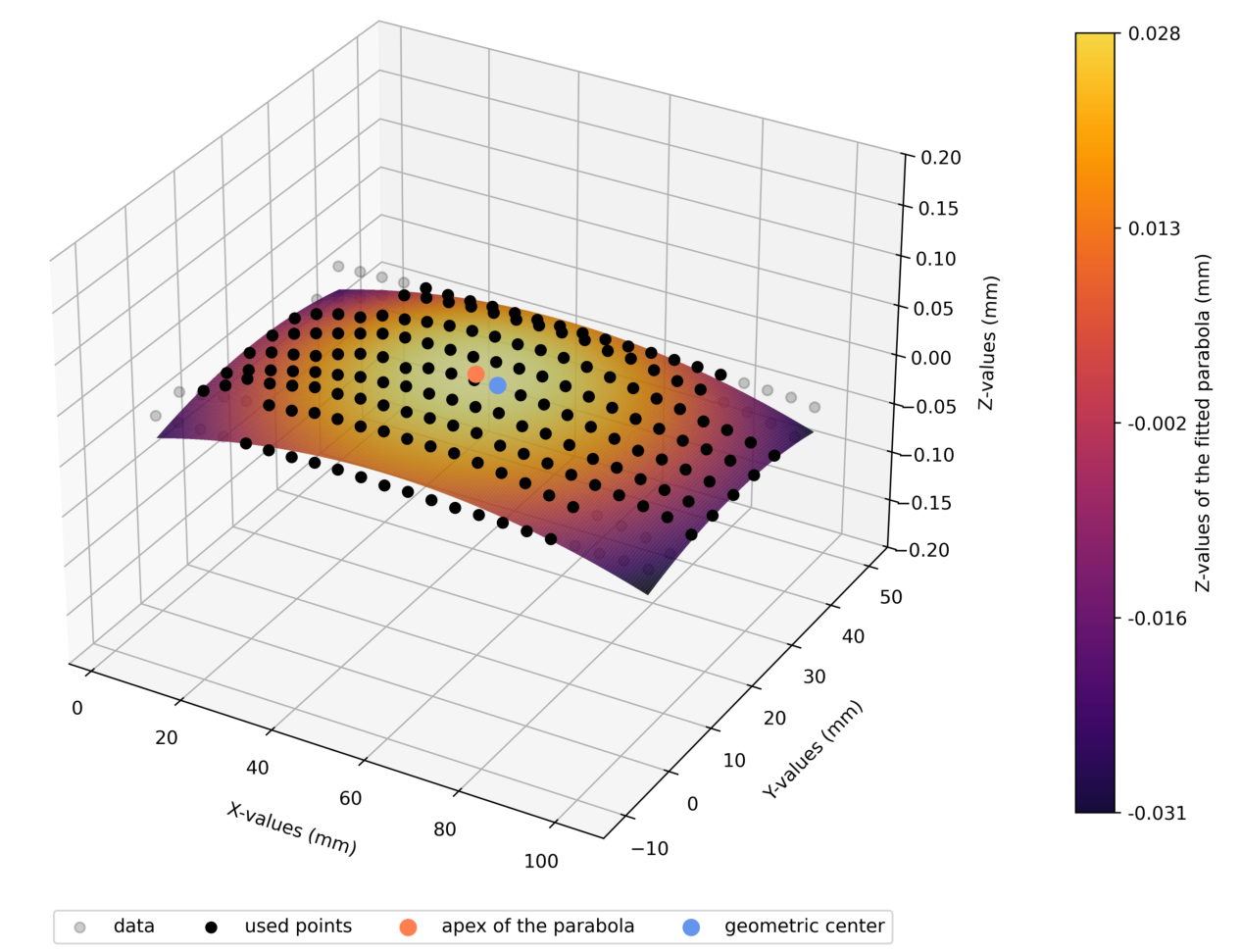}
\caption{Measured surface profile of the SPO sample. Finite-element simulations show that the deformation introduced by the measurement probe is only a few nanometers and therefore negligible.}
\label{fig:SPOshape}
\end{figure}

The measured surface profile is shown in Figure~\ref{fig:SPOshape}. A characteristic pillow-shaped curvature is clearly visible, with a height difference of approximately $50\,\upmu$m between the central region and both ends of the sample. Assuming an effective length of around 100\,mm, the maximum local surface inclination is

\[
\delta =
\frac{2\times50\,\mathrm{\upmu m}}{100\,\mathrm{mm}}
=
1\,\mathrm{mrad}
\approx
0.06^\circ .
\]

Although this value is comparable to the smallest grazing angles investigated, it is insufficient to explain the observed discrepancy. Moreover, the measured profile is well approximated by two parabolic functions in the longitudinal and transverse directions, resulting in an even flatter central region that dominates the scattering measurement. For a beam diameter of approximately $50$--$100\,\upmu$m behind the final slit pair, the footprint on the mirror extends over only 2.5--5\,cm at the smallest grazing angles investigated. Thus, only an even smaller portion of the maximum curvature is probed and the possible effect of mimicking a larger grazing angle is limited to well below 0.1\,mrad. 

These findings led to two immediate consequences. First, an independent and reliable method for the absolute calibration of the grazing angle had to be developed, as described in Section~\ref{ssec:inc}. Second, the scattering measurements had to be repeated with a demonstrably flat mirror sample. The commissioning run for these measurements is sketched in Section~\ref{ssec:DTU}.


\subsection{Determination of grazing angle}
\label{ssec:inc}

Several improvements were implemented to establish a reliable and reproducible determination of the proton grazing angle. In the previous Van de Graaff setup, the angle was obtained by aligning a laser beam coaxially with the proton beam and measuring the separation between the direct and the reflected laser spot on the detector plane. Due to the beam chopper installed at the ion implanter, however, the proton beam no longer follows a straight optical path, making this approach impossible.

\begin{figure}[ht]
\centering
\includegraphics[width=.6\textwidth]{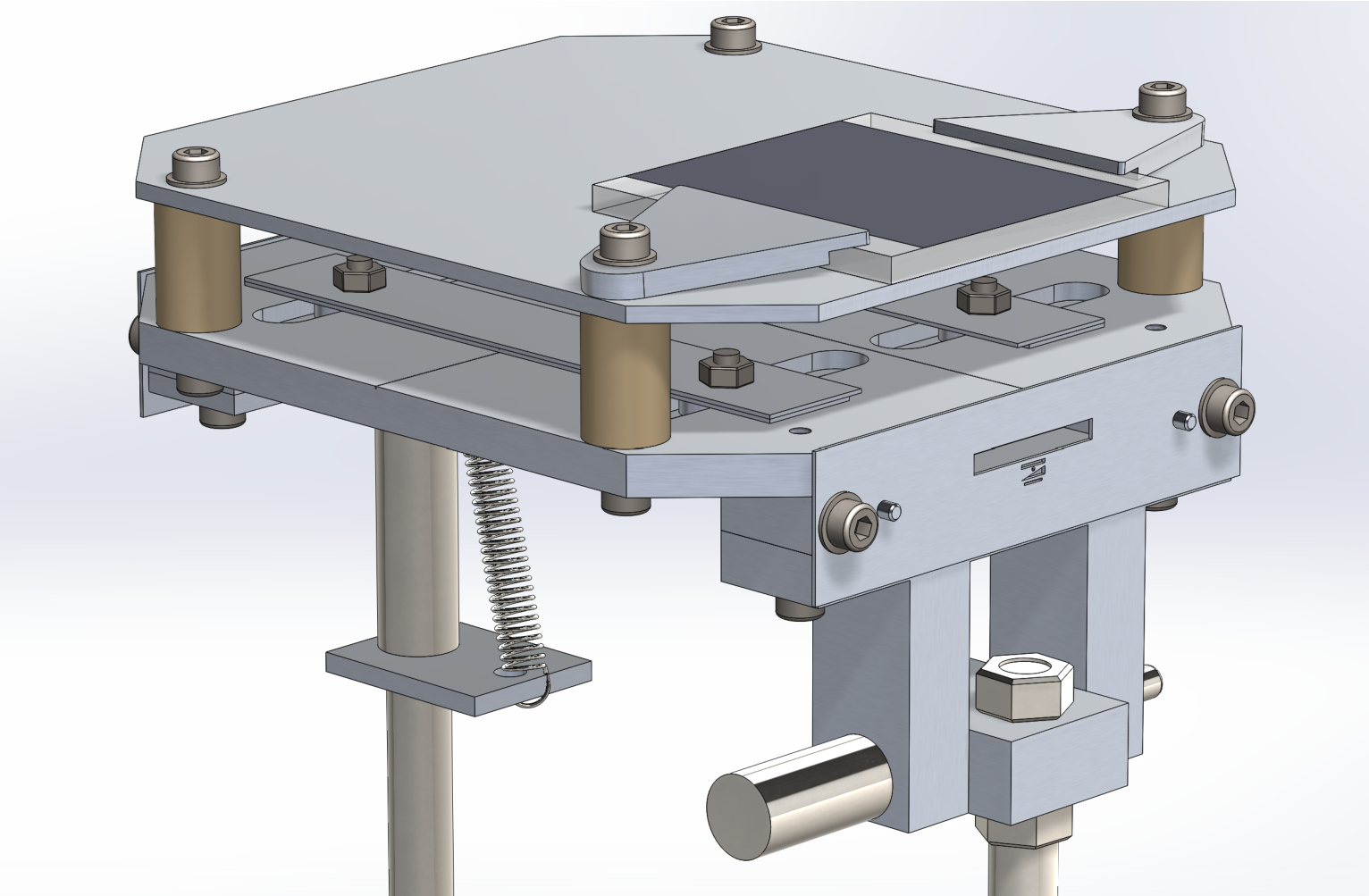}
\caption{CAD model of the upgraded target mount. Precision slit masks are mounted directly in front of and behind the target. An optical mirror fixed to the top of the mount is used for the laser-based monitoring of the target orientation.}
\label{fig:mount}
\end{figure}

\begin{figure}[ht]
\centering
\begin{tabular}{lr}
\includegraphics[width=.45\textwidth]{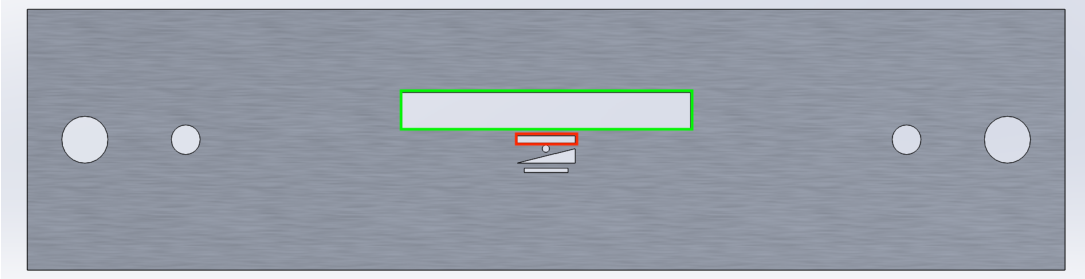} &
\includegraphics[width=.45\textwidth]{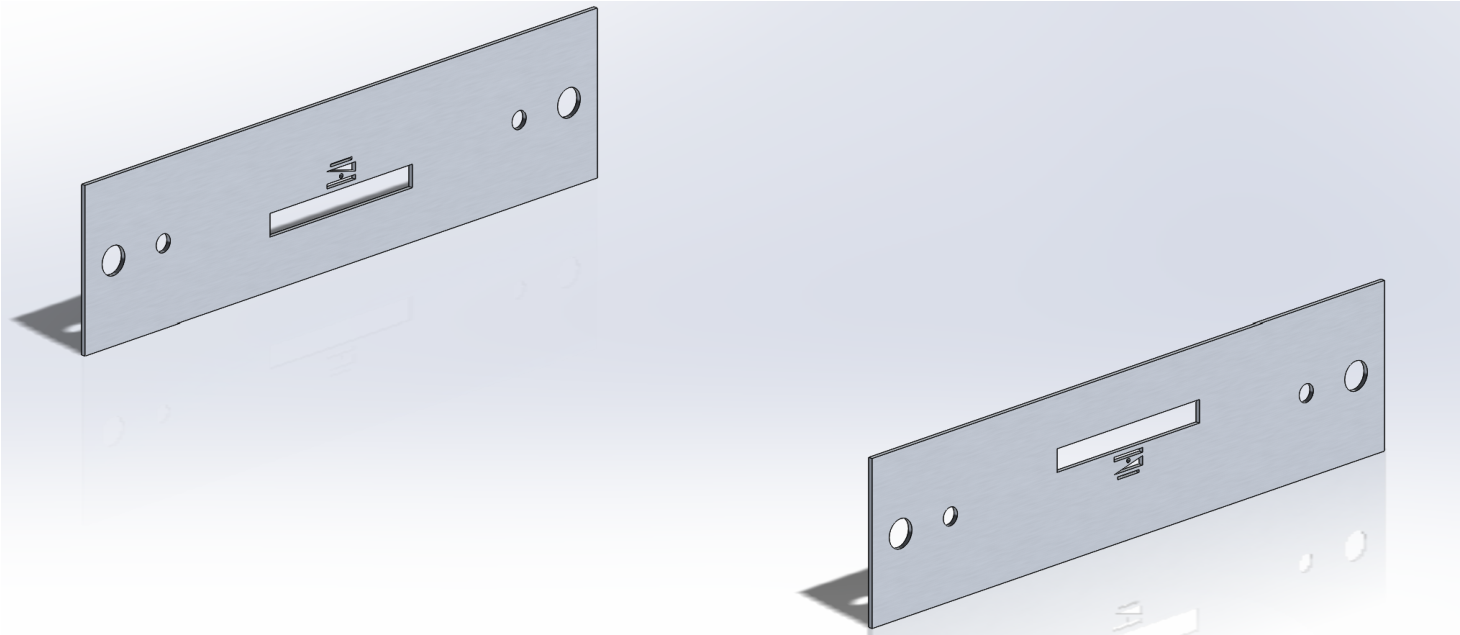}
\end{tabular}
\caption{Left: Precision slit mask with the relevant apertures highlighted. Right: Relative orientation of the two slit masks after installation.}
\label{fig:slits}
\end{figure}

Instead, two precision-manufactured slit masks were installed directly in front of and behind the target on the target mount. The distances between the slit masks and the mirror surface were measured with micrometer accuracy, comparable to the positioning accuracy of the linear manipulators supporting the target mount. The upgraded target mount is shown in Figure~\ref{fig:mount}, while Figure~\ref{fig:slits} presents close-ups on the slit mask and its configuration on the target mount.

The manipulator readings were calibrated by recording the shadows of the upper and lower slit edges on the detector and relating them to the mirror position. The resulting calibration was found to be consistent with the lower boundary of the scattering distribution measured for a flat target (see Section~\ref{ssec:DTU}).

\begin{figure}[ht]
\centering
\begin{tabular}{lr}
\includegraphics[width=.55\textwidth]{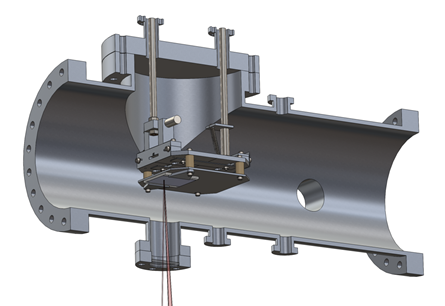} &
\includegraphics[width=.4\textwidth]{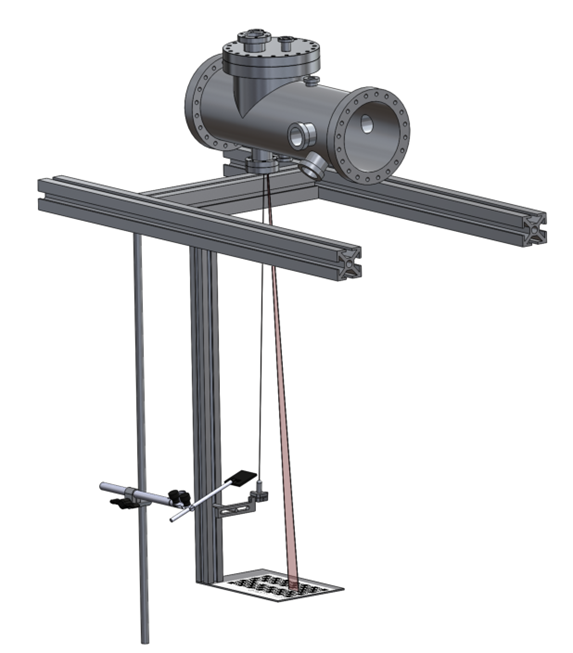}
\end{tabular}
\caption{Left: CAD cross section of the beamline showing the optical laser path. Right: Overview of the external laser arrangement used to monitor the target orientation.}
\label{fig:beamline}
\end{figure}

To provide an independent and rapid verification of the angular setting between successive measurements, an optical laser system was implemented. A laser beam enters the vacuum chamber through an optical window, is reflected by the mirror mounted on the target holder, and exits through the same window. The optical arrangement is illustrated in Figure~\ref{fig:beamline}. The reflected laser spot is recorded on a calibration target below the beamline. From the displacement of the spot and the known propagation distance, changes in the target orientation can be determined with high precision. Image analysis software automatically identifies calibration markers, performs the pixel-to-distance calibration, and corrects geometric distortions. The measured angular changes agree with those derived from the manipulator readings within the experimental uncertainty.

\subsection{Setup commissioning with flat multilayer target}
\label{ssec:DTU}

The upgraded target mount and the newly implemented methods for determining the absolute and relative grazing angle were commissioned during a first measurement campaign in October 2025. To exclude target-specific effects as the origin of the observed discrepancy, a flat multilayer X-ray mirror prototype manufactured at the Department of Space Research and Technology at the Technical University of Denmark (DTU Space) was used.

\begin{figure}[ht]
  \begin{center}
      \includegraphics[width=.5\textwidth]{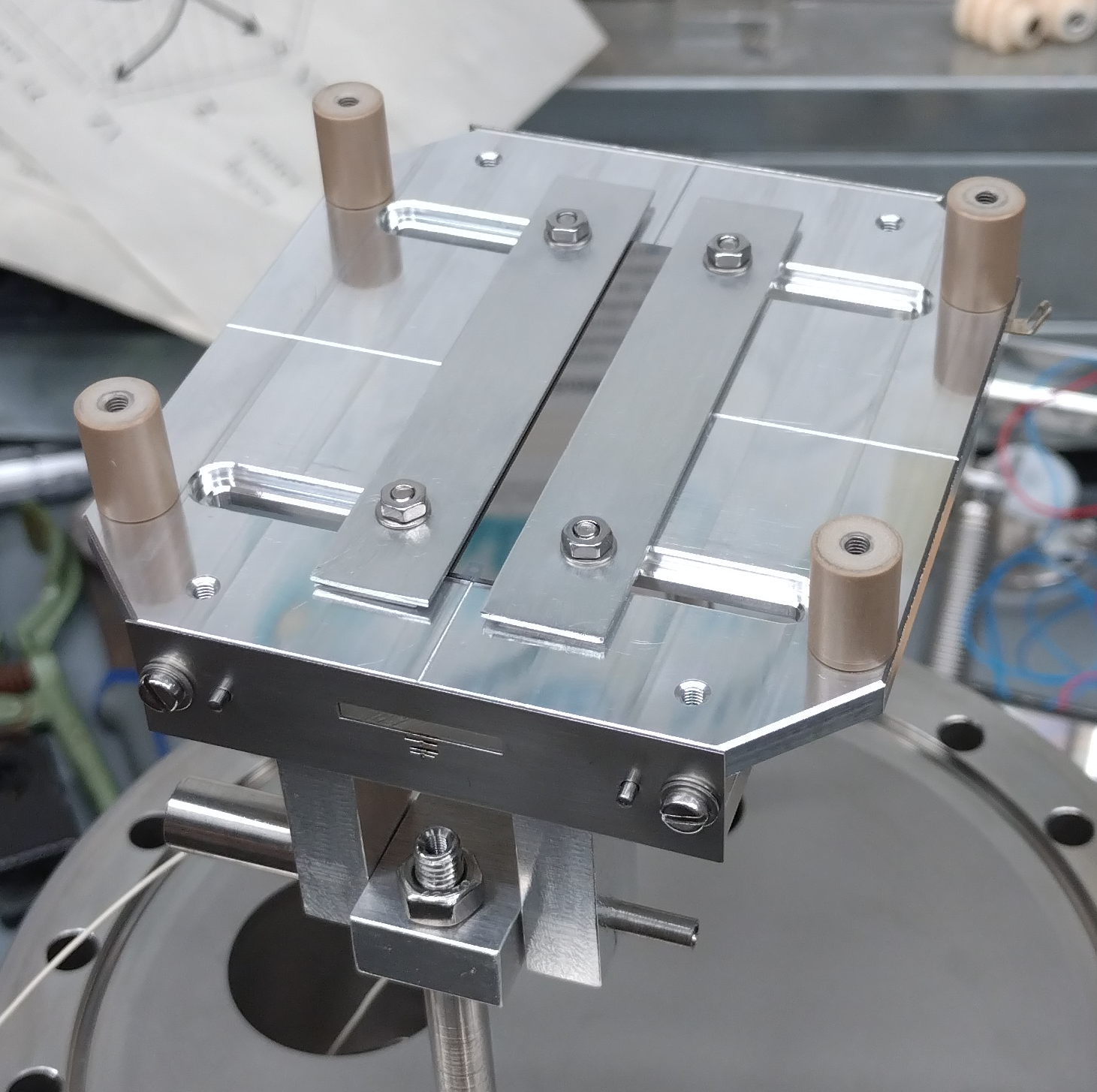}
  \end{center}
  \caption{\label{fig:dtu_mirror}
  The upgraded target mount with the novel slit mask on the frontside positioned using high-precision conical pins. The mounted target is a flat multilayer X-ray mirror from DTU Space. The upper metal plate with the optical mirror for the laser measurement is removed for visibility.}
\end{figure}

The mirror is based on the multilayer technology developed for \textit{NuSTAR} \cite{Harrison2013}, although with a significantly larger number of coating layers. It consists of a 0.8\,mm thick silicon substrate coated with 50 W/SiC bilayers. While the thickness is comparable to that of the SPO samples, the mirror dimensions are considerably smaller, with a length of 70\,mm and a width of only 10\,mm. Figure~\ref{fig:dtu_mirror} shows the sample mounted on the upgraded holder. Due to its narrow width, careful alignment of the proton beam with the mirror surface was required to avoid shadowing by the mounting structure.

\begin{figure}[ht]
  \begin{center}
    \includegraphics[width=\textwidth]{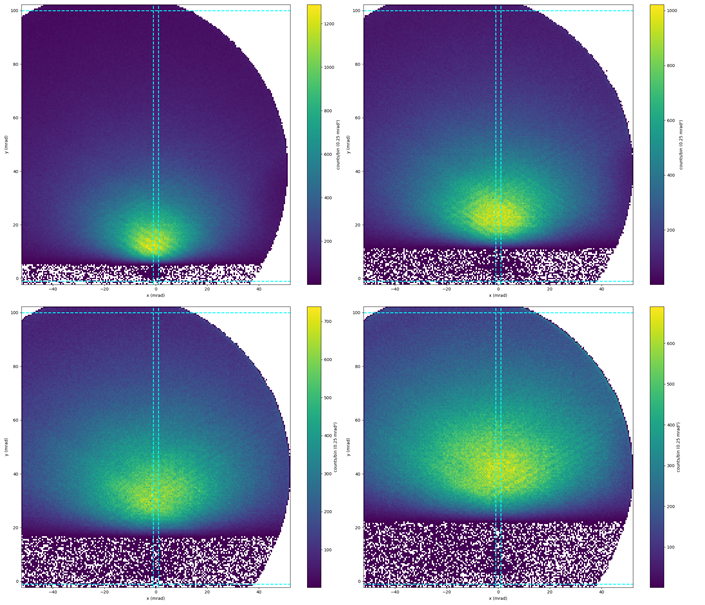}
  \end{center}
  \caption{\label{fig:distributions}
  Preliminary results of the scattering distributions at 20\,keV in arbitrary units with the flat multilayer X-ray mirror target from DTU Space. The region enclosed by the dashed lines was used to analyze the scattering profiles.}
\end{figure}

Scattering measurements were performed at proton energies of 20\,keV and 50\,keV for grazing angles between approximately 5\,mrad and 20\,mrad. The preliminary scattering distributions obtained at 20\,keV are shown in Figure~\ref{fig:distributions}. Since the detector had not yet been fully calibrated, a hexagonal response pattern is visible and the data are shown in arbitrary units without normalization to the incident beam current. The region enclosed by the dashed lines was used for a first comparison with the SPO measurements.

Despite the substantially improved determination of the grazing angle, the preliminary results again show a systematic shift of the scattering maximum towards larger angles. This observation suggests that the discrepancy with the Remizovich model may not originate from the SPO samples themselves. However, measurements with the fully calibrated detector are required before firm conclusions can be drawn.


\section{SOFT PROTON TRANSMISSION TEST}
\label{sec:trans}

Besides grazing-incidence scattering experiments and irradiation campaigns for radiation-hardness studies, a third application of the beamline is the investigation of soft proton transmission through thin foils. Of particular interest are proton straggling, energy loss, and charge exchange in X-ray filters, as these quantities are required for the assessment of future X-ray instruments. Thin entrance filters are commonly placed in the optical path of X-ray telescopes to suppress infrared, visible, and ultraviolet light, low-energy particles, as well as contamination, while maintaining a high transmission for X-rays.

\begin{figure}[ht]
\centering
\includegraphics[width=.5\textwidth]{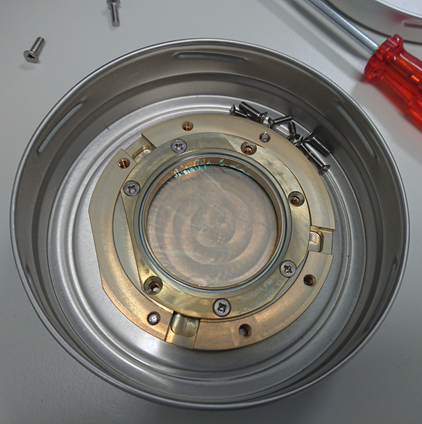}
\caption{The flight spare \textit{eROSITA} X-ray filter used for the proton transmission measurements.}
\label{fig:eROSITA}
\end{figure}

To evaluate the suitability of the implanter setup for such measurements, a first commissioning experiment was carried out in August 2025. The investigated sample was a flight spare filter of the \textit{eROSITA} instrument, manufactured by MOXTEK\footnote{\url{https://moxtek.com/}} and kindly provided by the Max Planck Institute for Extraterrestrial Physics (MPE)\footnote{\url{https://www.mpe.mpg.de/}}. The filter consists of a 192.4\,nm thick polyimide film mounted in the original \textit{eROSITA} filter frame with a free aperture of 42\,mm (see Figure~\ref{fig:eROSITA}).

The filter was mounted on the same flange used for the scattering targets, at a distance of approximately 84\,cm from the MCP detector. A push--pull manipulator allowed the filter to be inserted into and removed from the proton beam without venting the beamline. Due to the fragility of the filter, particular care was taken during installation, and during handling and vacuum operations it was kept in a protected parking position between two metal plates.

To minimize distortions of the transmitted proton distribution, the Earth's magnetic field was compensated by external Helmholtz coils over the flight path between the filter and the detector; likewise as for the grazing-incidence scattering. Test measurements were performed at proton energies of 20\,keV, 35\,keV, and 50\,keV.

The preliminary results are in qualitative agreement with SRIM/TRIM simulations. The measured distributions appear slightly broader and exhibit a somewhat larger energy loss than predicted. One possible explanation is the presence of surface contamination, for example through adsorbed water layers. A quantitative analysis, however, was hampered by an additional background contribution originating from ions that were backscattered from the inner walls of the beamline. This effect became particularly pronounced at lower proton energies, where multiple scattering inside the filter increases the beam divergence.

\begin{figure}[ht]
\centering
\includegraphics[width=.5\textwidth]{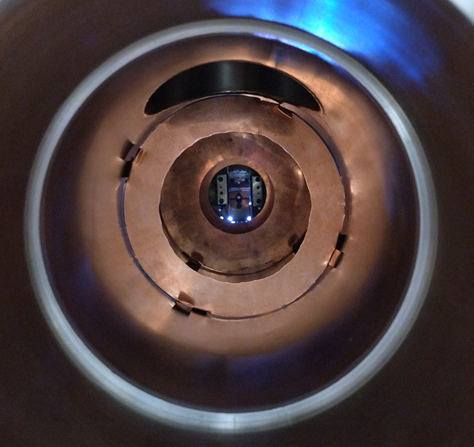}
\caption{Copper apertures installed inside the beamline to suppress background from protons backscattered at the inner beamline walls. The photograph is taken from the detector side toward the target position.}
\label{fig:antiscat}
\end{figure}

To suppress this background, three copper apertures with progressively increasing inner diameters were installed downstream of the target position. Their purpose is to intercept protons scattered from the beamline walls before they reach the detector. The modified beamline is shown in Figure~\ref{fig:antiscat}.

The upgraded beamline will be used in a dedicated transmission campaign to obtain quantitative measurements of proton energy loss, angular straggling, and charge-exchange fractions for the \textit{eROSITA} filter. These measurements will also provide a direct validation of TRIM and Geant4 simulations in the energy range 20--50\,keV.


\section{CONCLUSIONS}
\label{sec:concl}

A new low-energy soft proton scattering facility has been established at the ion implanter of the IKF, extending experimental investigations to proton energies between approximately 10\,keV and 60\,keV. During the first measurement campaign with an SPO sample, systematic deviations from the Remizovich model were observed at the smallest grazing angles.

A detailed investigation showed that neither image-charge effects nor the intrinsic curvature of the SPO sample can fully account for the observed shift of the scattering distributions. This motivated the development of an improved target mount together with new methods for the absolute and relative determination of the grazing angle. Commissioning measurements using a flat multilayer mirror from DTU Space indicate that the systematic shift persists, suggesting that the discrepancy may be intrinsic rather than caused by the target geometry. However, further measurements with a fully calibrated detector are required before definitive conclusions can be drawn.

The suitability of the new setup for transmission experiments was demonstrated using a flight spare \textit{eROSITA} X-ray filter. The commissioning measurements revealed a previously underestimated background contribution from ions backscattered at the inner beamline walls. To suppress this effect, a system of copper apertures has been installed downstream of the target, improving both transmission and scattering measurements.

The immediate next steps are the full calibration of the hexanode MCP detector, including timing optimization and sensitivity flat-fielding, followed by dedicated scattering campaigns with both SPO and flat multilayer mirror targets. Complementary transmission measurements of the \textit{eROSITA} filter will provide quantitative data on angular straggling, energy loss, and charge exchange for comparison with TRIM and Geant4 simulations. In addition, quantitative measurements of the charge state after scattering and transmission are planned. Since preliminary results indicate that most low-energy protons undergo charge exchange and emerge as neutral hydrogen atoms, these measurements will provide important input for assessing the effectiveness and design requirements of magnetic proton diverters for future X-ray observatories.

\bibliography{2026-SPIE-SP} 
\bibliographystyle{spiebib} 

\end{document}